\PassOptionsToPackage{table,xcdraw}{xcolor}
\documentclass[12pt,a4paper]{article}
\pdfoutput=1

\usepackage{geometry}
\usepackage{comment}
\usepackage[numbers,sort&compress]{natbib}
\usepackage{amsmath}
\usepackage{amssymb}
\usepackage[dvips]{graphicx}
\usepackage{pstricks}
\usepackage{bm}
\usepackage{pbox}
\usepackage{placeins}
\usepackage{graphicx}
\usepackage{caption}
\usepackage{subcaption}
\usepackage[T1]{fontenc}
\usepackage{footnote}
\usepackage{pdfpages}
\usepackage{hhline}
\usepackage{multirow}
\usepackage{enumitem}
\usepackage{slashed}
\usepackage[nottoc,notlot,notlof]{tocbibind}
\usepackage[title,titletoc]{appendix}
\usepackage{dsfont}
\usepackage{pifont}
\allowdisplaybreaks
\usepackage{cases}
\usepackage{tikz}
\usepackage[symbol]{footmisc}
\usepackage{physics}
\usepackage{soul,listings}
\usepackage{booktabs}
\usepackage{longtable}

\usepackage{tabularx}
\newcolumntype{C}{>{\centering\arraybackslash}X}
\newcolumntype{L}{>{\centering\arraybackslash}m{2.1cm}}
\newcolumntype{N}{>{\centering\arraybackslash}m{1.3cm}}

\def\equationautorefname~#1\null{Eq.\,(#1)\null}

\definecolor{MyDarkBlue}{rgb}{0.1, 0.1, 0.8}
\definecolor{MyLightBlue}{rgb}{0.22,0.51,0.9}
\definecolor{MyGreen}{rgb}{0.0, 0.5, 0.0}
\definecolor{BrickRed}{rgb}{0.8, 0.25, 0.33}
\usepackage[colorlinks=true,linkcolor=blue,citecolor=MyDarkBlue,
urlcolor=MyLightBlue,bookmarksnumbered=true,bookmarksopen]{hyperref}
\hypersetup{colorlinks, citecolor=blue,linkcolor=blue, urlcolor=MyLightBlue}

\begin{document}
\vspace*{-0.2in}
\begin{flushright}
ZTF-EP-26-10
\end{flushright}
\vspace{0.5cm}
\begin{center}
{\Large\bf   
Predictive Non-Minimal $SU(5)$ GUT
}
\end{center}

\vspace{0.5cm}
\renewcommand{\thefootnote}{\fnsymbol{footnote}}
\begin{center}
{\large
{}~\textbf{Ilja Dor\v{s}ner$^{1}$}\footnote[1]{ E-mail: \textcolor{MyLightBlue}{dorsner@phy.hr}}, 
{}~\textbf{Mijo Matkovi\'c$^{1,2}$}\footnote[3]{ E-mail: \textcolor{MyLightBlue}{mijo.matkovic@fesb.hr}}, 
{}~\textbf{Shaikh Saad$^3$}\footnote[2]{E-mail:  
\textcolor{MyLightBlue}{shaikh.saad@ijs.si}
}
}
\vspace{0.5cm}

{\em $^1$ University of Zagreb Faculty of Science, Department of Physics, Bijeni\v cka cesta 32, 10000 Zagreb, Croatia}
\\
{\em $^2$ University of Split, Faculty of Electrical Engineering, Mechanical Engineering and\\ Naval Architecture, Ru\dj era Bo\v{s}kovi\'{c}a 32, 21000 Split, Croatia}
\\
{\em $^3$Jožef  Stefan Institute, Jamova 39,
  1000 Ljubljana, Slovenia}
\end{center}

\renewcommand{\thefootnote}{\arabic{footnote}}
\setcounter{footnote}{0}
\thispagestyle{empty}

%%%%%%%%%%%%%%%%%%%%%%%%%%%%%
\begin{abstract}
The breaking of $SU(5)$ down to the Standard Model gauge group can be implemented by any non-trivial self-conjugate scalar representation with a suitable vacuum expectation value. The Georgi-Glashow model accomplishes this breaking with a 24-dimensional representation. That choice, although the most economical one, fails to unify the gauge coupling constants. Moreover, the subsequent breaking of the Standard Model down to $SU(3) \times U(1)_\mathrm{em}$ with a $5$-dimensional representation fails to simultaneously provide viable masses for the down-type quarks and the charged leptons. We show that the substitution of the $24$-dimensional representation with a $75$-dimensional representation  fixes the gauge coupling unification issue outrightly. We furthermore pin down, by computing the full mass spectrum of the multiplets in $75$-dimensional representation, the range of the unification scale $m_{\rm GUT}$ to be $10^{15}\,\text{GeV} \lesssim m_{\rm GUT} \lesssim 10^{16}\,\text{GeV}$. This  unification window is partially excluded by Super-Kamiokande data and will be experimentally  accessible at Hyper-Kamiokande. The second issue with the  Georgi-Glashow model can be addressed, for example, with vectorlike fermions. The $24$-dimensional scenario admits three different types of vectorlike fermions that can properly account for the mismatch between the down-type quark and the charged lepton masses. The $75$-dimensional scenario, on the other hand, offers enhanced predictivity by allowing only a unique vectorlike addition of  $10_F + \overline{10}_F$ that restores realistic charged fermion masses  and modestly pushes the upper limit on the unification scale to $m_{\rm GUT} \lesssim 3\times10^{16}\,\text{GeV}$.  The proposed framework with $75$-dimensional scalar representation can thus serve as a  phenomenologically viable alternative to the standard Georgi-Glashow symmetry breaking paradigm.
\end{abstract}
%%%%%%%%%%%%%%%%%%%%%%%%%%%%%
\newpage
\setcounter{footnote}{0}

%%%%%%%%%%%%%%%%%%%%%%%%%%%%%
\section{Introduction}
%%%%%%%%%%%%%%%%%%%%%%%%%%%%%
The simplest possible group that can unify gauge interactions of the Standard Model (SM) is $SU(5)$. Since both $SU(5)$ and the SM gauge group $SU(3) \times SU(2) \times U(1)$ are of the same rank, there is only one possible symmetry breaking direction that can take the former into the latter. One might say that this property is one of the most appealing features of $SU(5)$. 

The dimensionality of the irreducible representation that implements the symmetry breaking chain $SU(5) \rightarrow SU(3) \times SU(2) \times U(1)$, on the other hand, is not uniquely determined. Namely, any non-trivial self-conjugate scalar representation with appropriate vacuum expectation value (VEV) might provide viable symmetry breaking of $SU(5)$ down to $SU(3) \times SU(2) \times U(1)$. The first three such irreducible representations, in the tensorial language of $SU(5)$, are $24^i_j$~\cite{Georgi:1974sy}, $75^{ij}_{kl}=-75^{ji}_{kl}=-75^{ij}_{lk}$~\cite{Hubsch:1984pg,Hubsch:1984qi}, and $200^{ij}_{kl}=200^{ji}_{kl}=200^{ij}_{lk}$~\cite{Hubsch:1984zi}, where $i,j,k,l=1,\ldots,5$ are group indices. The dimension of each of these representations is self-evident, whereas contraction of any upper index with any lower index, within a given irreducible representation, always yields a zero. We opt to denote scalar representations with their dimensionality throughout this manuscript to simplify notation.

The original proposal~\cite{Georgi:1974sy} of $SU(5)$ unification utilizes a 24-dimensional scalar representation to break $SU(5)$ down to $SU(3) \times SU(2) \times U(1)$. This is certainly the most economical setup, but, due to its minimality, the Georgi-Glashow model fails to provide gauge coupling unification if the subsequent breaking of $SU(3) \times SU(2) \times U(1)$ down to $SU(3) \times U(1)_\mathrm{em}$ is accomplished via an electroweak VEV of a 5-dimensional scalar representation. It also fails to simultaneously accommodate phenomenologically viable masses of the down-type quarks and the charged leptons. 

We are thus prompted to analyze what happens when a $24^i_j$ of the minimal Georgi-Glashow setup is replaced with a $75^{ij}_{lk}$. The group theoretical technicalities of the symmetry breaking of $SU(5)$ with $75^{ij}_{lk}$ have already been thoroughly investigated, but the phenomenological aspects such as gauge coupling unification, proton decay, and viable fermion mass generation have not. We plan to rectify this in a the present study. 

We find that, unlike the $24$-dimensional scenario of the Georgi-Glashow model, the $75$-dimensional one successfully achieves gauge coupling unification.
The predicted upper limit on unification scale comes out to be  $\lesssim 10^{16}$\,GeV. This setup thus offers direct testability in upcoming proton decay experiments such as DUNE and Hyper-Kamiokande. We also find that, if vector-like fermions are employed to correct unrealistic mass relations between the down-type quarks and the charged leptons, the $75$-dimensional scenario can only be extended with one particular vectorlike fermion type, unlike the $24$-dimensional scenario that allows for use of three different types. 

The manuscript is organized as follows. In Sec.\ \ref{sec:one} we investigate phenomenology of the scenario when a $24$-dimensional scalar representation within the original Georgi-Glashow $SU(5)$ proposal is simply replaced with a $75$-dimensional one. In the process we also double-check existing results on the subject in the literature. Sec.\ \ref{sec:two} contains analysis of an $SU(5)$ model with a $75$-dimensional scalar representation that can simultaneously yield gauge coupling unification and accommodate viable masses of all the SM fermions. There we demonstrate that the  $75$-dimensional scalar setup can only use one particular type of vectorlike fermions to break unwanted  degeneracy between the masses of the down-type quarks and the charged leptons. We briefly conclude in Sec.\ \ref{sec:three}.

%%%%%%%%%%%%%%%%%%%%%%%%%%%%%
\section{$75$ instead of $24$}
%%%%%%%%%%%%%%%%%%%%%%%%%%%%%
\label{sec:one}
We first study  phenomenology of the Georgi-Glashow~\cite{Georgi:1974sy} $SU(5)$ setup when a  24-dimensional scalar representation is simply replaced with a 75-dimensional~\cite{Hubsch:1984pg,Hubsch:1984qi} one. We demonstrate that this replacement, i.e., $24^i_j \rightarrow 75^{ij}_{kl}$, offers gauge  coupling unification and sets an upper bound on the  scale of unification that is significantly lower than the Planck scale. Note that we are solely interested in renormalizable model(s).
We also stress that the fermions of this particular scenario  are exactly those of the Georgi-Glashow model and comprise  $\overline{5}^A_{F i}$ and $10^{ijB}_{F}=-10^{jiB}_{F}$, where $A,B=1,2,3$ are family indices,  $i,j=1,\ldots,5$ are $SU(5)$ indices, and $F$ denotes fermionic nature of representation. Both the scalars and the fermions of the scenario under consideration are summarized in unshaded part of Table \ref{tab:ParticleContent}.

\begin{table}[h!]
    \centering
    \renewcommand{\arraystretch}{0.9}
    \begin{tabularx}{\textwidth}{|L|C|C|C|N|N|}
        \hline
         & $SU(5)$ & $J$(SM) & $(b^J_1, b^J_2, b^J_3)$ & $b^J_{12}$ & $b^J_{23}$ \\
        \hline\hline
        \multirow{11}{*}{Scalars}
         & \multirow{2}{*}{$\Lambda^i = 5^i$} & $\Lambda_2 (1, 2, \frac{1}{2})$ & $(\frac{1}{10}, \frac{1}{6}, 0)$ & $-\frac{1}{15}$ & $\frac{1}{6}$ \\
         & & $\Lambda_3 (3, 1, -\frac{1}{3})$ & $(\frac{1}{15}, 0, \frac{1}{6})$ & $\frac{1}{15}$ & $-\frac{1}{6}$ \\
        \cline{2-6}
         & \multirow{9}{*}{$\Phi^{ij}_{kl} = 75^{ij}_{kl}$}
         & $\Phi_1(1, 1, 0)$ & $(0, 0, 0)$ & $0$ & $0$ \\
         & & $\Phi_{8,1}(8, 1, 0)$ & $(0, 0, \frac{1}{2})$ & $0$ & $-\frac{1}{2}$ \\
         & & $\Phi_{8,3}(8, 3, 0)$ & $(0, \frac{8}{3}, \frac{3}{2})$ & $-\frac{8}{3}$ & $\frac{7}{6}$ \\
         & & $\Phi_{3,2}(3, 2, -\frac{5}{6})$ & $(\frac{5}{12}, \frac{1}{4}, \frac{1}{6})$ & $\frac{1}{6}$ & $\frac{1}{12}$ \\
         & & $\Phi_{\overline{3},2}(\overline{3}, 2, \frac{5}{6})$ & $(\frac{5}{12}, \frac{1}{4}, \frac{1}{6})$ & $\frac{1}{6}$ & $\frac{1}{12}$ \\
         & & $\Phi_{\overline{3},1}(\overline{3}, 1, -\frac{5}{3})$ & $(\frac{5}{6}, 0, \frac{1}{12})$ & $\frac{5}{6}$ & $-\frac{1}{12}$ \\
         & & $\Phi_{3,1}(3, 1, \frac{5}{3})$ & $(\frac{5}{6}, 0, \frac{1}{12})$ & $\frac{5}{6}$ & $-\frac{1}{12}$ \\
         & & $\Phi_{\overline{6},2}(\overline{6}, 2, -\frac{5}{6})$ & $(\frac{5}{6}, \frac{1}{2}, \frac{5}{6})$ & $\frac{1}{3}$ & $-\frac{1}{3}$ \\
         & & $\Phi_{6,2}(6, 2, \frac{5}{6})$ & $(\frac{5}{6}, \frac{1}{2}, \frac{5}{6})$ & $\frac{1}{3}$ & $-\frac{1}{3}$ \\
        \hline
        \hline
        \multirow{11}{*}{Fermions}
         & \multirow{2}{*}{$\overline{5}^A_{Fi}$} & $L_a(1, 2, -\frac{1}{2})$ & $(\frac{3}{5}, 1, 0)$ & $-\frac{2}{5}$ & $1$ \\
         & & $d_a^c(\overline{3}, 1, \frac{1}{3})$ & $(\frac{2}{5}, 0, 1)$ & $\frac{2}{5}$ & $-1$ \\
        \cline{2-6}
         & \multirow{3}{*}{$10^{ijA}_{F}$} & $q_a(3, 2, \frac{1}{6})$ & $(\frac{1}{5}, 3, 2)$ & $-\frac{14}{5}$ & $1$ \\
         & & $u_a^c(\overline{3}, 1, -\frac{2}{3})$ & $(\frac{8}{5}, 0, 1)$ & $\frac{8}{5}$ & $-1$ \\
         & & $e_a^c(1, 1, 1)$ & $(\frac{6}{5}, 0, 0)$ & $\frac{6}{5}$ & $0$ \\
        \cline{2-6}
         % Shaded block 1 (10_F)
         & \cellcolor{gray!20} & \cellcolor{gray!20} $e^c_4(1, 1, 1)$ & \cellcolor{gray!20} $(\frac{2}{5}, 0, 0)$ & \cellcolor{gray!20} $\frac{6}{15}$ & \cellcolor{gray!20} $0$ \\
         & \cellcolor{gray!20} & \cellcolor{gray!20} $u^c_4(\overline{3}, 1, -\frac{2}{3})$ & \cellcolor{gray!20} $(\frac{8}{15}, 0, \frac{1}{3})$ & \cellcolor{gray!20} $\frac{8}{15}$ & \cellcolor{gray!20} $-\frac{1}{3}$ \\
         & \cellcolor{gray!20} \multirow{-3}{*}{$10^{ij4}_{F}$} & \cellcolor{gray!20} $q_4(3, 2, \frac{1}{6})$ & \cellcolor{gray!20} $(\frac{1}{15}, 1, \frac{2}{3})$ & \cellcolor{gray!20} $-\frac{14}{5}$ & \cellcolor{gray!20} \strut $\frac{1}{3}$ \\
        \cline{2-6}
         % Shaded block 2 (\overline{10}_F)
         & \cellcolor{gray!20} & \cellcolor{gray!20}\strut$E(1, 1, -1)$ & \cellcolor{gray!20}$(\frac{2}{5}, 0, 0)$ & \cellcolor{gray!20}\strut$\frac{6}{15}$ & \cellcolor{gray!20}$0$ \\
         & \cellcolor{gray!20} & \cellcolor{gray!20} $U(3, 1, \frac{2}{3})$ & \cellcolor{gray!20} $(\frac{8}{15}, 0, \frac{1}{3})$ & \cellcolor{gray!20} $\frac{8}{15}$ & \cellcolor{gray!20} $-\frac{1}{3}$ \\
         & \cellcolor{gray!20}\multirow{-3}{*}{$\overline{10}_{Fij}$} & \cellcolor{gray!20} $Q^c(\overline{3}, 2, -\frac{1}{6})$ & \cellcolor{gray!20} $(\frac{1}{15}, 1, \frac{2}{3})$ & \cellcolor{gray!20} $-\frac{14}{5}$ & \cellcolor{gray!20} $\frac{1}{3}$ \\ 
        \hline
    \end{tabularx}
    \caption{Scalars and fermions of two $SU(5)$ scenarios that are featured in the manuscript with the explicit SM decomposition and associated $\beta$-coefficients. Here, $i,j,k,l=1,\ldots,5$ are $SU(5)$ indices and $A=1,2,3$ is a flavor index.  }
    \label{tab:ParticleContent}
\end{table}

%%%%%%%%%%%%%%%%%%%%%%%%%%%%%%%%%%%%%%%%%%%%%%%%
\subsection{Scalar mass spectrum}
The symmetry properties of a 75-dimensional scalar representation are  $75^{ij}_{lk}=-75^{ji}_{lk}=-75^{ij}_{kl}$, with $75^{ij}_{ik}=0$, whereas the field decomposition under the SM gauge group is 
\begin{align}
    75=\Phi&=
\Phi_1(1,1,0) + \Phi_{8,1}(8,1,0) + \Phi_{8,3}(8,3,0)
\nonumber \\  &
+ \bigg\{ \Phi_{3,2}(3,2,-5/6)
+ \Phi_{\overline 6,2}(\overline 6,2,-5/6)
+  \Phi_{\overline 3,1}(\overline 3,1,-5/3)
+ \text{c.c.}  \bigg\},
\end{align}
where we also establish notation for the SM multiplets in $75^{ij}_{kl}$. 

The exact tensorial decomposition of $\Phi$, with properly normalized scalar fields, reads: 
\begin{align}
\label{eq:a}
  \Phi_1
    ={}& s\Big[
      \delta^\alpha_{\gamma}\delta^\beta_{\delta}
      -\delta^\alpha_{\delta}\delta^\beta_{\gamma}
      +3\left(\delta^a_{c}\delta^b_{d}
      -\delta^a_{d}\delta^b_{c}\right) 
      -\left(
        \delta^\alpha_{\gamma}\delta^b_{d}
        -\delta^\alpha_{\delta}\delta^b_{c}
        +\delta^\beta_{\delta}\delta^a_{c}
        -\delta^\beta_{\gamma}\delta^a_{d}
       \right)
    \Big],\\
\label{eq:b}    
  \Phi_{8,1}
    ={}& -\frac{1}{2}\Big[
      (2\delta^\alpha_{\gamma}-\delta^a_{c})
        \widetilde\Phi^\beta_{\delta}
      -(2\delta^\alpha_{\delta}-\delta^a_{d})
        \widetilde\Phi^\beta_{\gamma} 
      +(2\delta^\beta_{\delta}-\delta^b_{d})
        \widetilde\Phi^\alpha_{\gamma}
      -(2\delta^\beta_{\gamma}-\delta^b_{c})
        \widetilde\Phi^\alpha_{\delta}
    \Big],\\
\label{eq:c}    
  \Phi_{8,3}
    ={}& \widetilde\Phi^{\alpha b}_{\gamma d}
       + \widetilde\Phi^{\alpha b}_{c\delta}
       + \widetilde\Phi^{\beta a}_{\delta c}
       - \widetilde\Phi^{\beta a}_{\gamma d},\\
\label{eq:d}       
  \Phi_{3,2}
    ={}& -\frac{1}{2}\Big[
      (\delta^\alpha_{\gamma}-2\delta^a_{c})
        \widetilde\Phi^\beta_{d}
      -(\delta^\alpha_{\delta}-2\delta^a_{d})
        \widetilde\Phi^\beta_{c} 
      +(\delta^\beta_{\delta}-2\delta^b_{d})
        \widetilde\Phi^\alpha_{c}
      -(\delta^\beta_{\gamma}-2\delta^b_{c})
        \widetilde\Phi^\alpha_{d}
    \Big],\\
\label{eq:e}    
  \Phi_{\overline 6,2}
    ={}& \widetilde\Phi^{\alpha\beta}_{\gamma d}
       + \widetilde\Phi^{\alpha\beta}_{c\delta},\\
\label{eq:f}       
  \Phi_{\overline 3,1}
    ={}& \widetilde\Phi^{\alpha\beta}_{cd}.
\end{align}
Here, $\alpha, \beta, \gamma, \delta =1,2,3$, whereas $a,b,c,d=4,5$. Also, the reduced
tensors that are featured in Eqs.\ \eqref{eq:b} through \eqref{eq:f} are 
\begin{align}
  \widetilde\Phi^\alpha_{\gamma}
    &= \Phi^{\alpha a}_{\gamma a}
       -\frac{1}{3}\delta^\alpha_{\gamma}
        \Phi^{\beta a}_{\beta a},
        \quad
  \widetilde\Phi^\alpha_{c}
    = \Phi^{\alpha a}_{c a},
    \quad
  \widetilde\Phi^{\alpha\beta}_{cd}
    = \Phi^{\alpha\beta}_{cd},\\
  \widetilde\Phi^{\alpha b}_{\gamma d}
    &= \Phi^{\alpha b}_{\gamma d}
       -\frac{1}{3}\delta^\alpha_{\gamma}
        \Phi^{\beta b}_{\beta d}
       -\frac{1}{2}\delta^b_{d}
        \Phi^{\alpha a}_{\gamma a}
       +\frac{1}{6}\delta^\alpha_{\gamma}\delta^b_{d}
        \Phi^{\beta a}_{\beta a},\\
  \widetilde\Phi^{\alpha\beta}_{\gamma d}
    &= \Phi^{\alpha\beta}_{\gamma d}
       -\frac{1}{2}\left(
        \delta^\alpha_{\gamma}\Phi^{\beta\delta}_{d\delta}
        -\delta^\beta_{\gamma}\Phi^{\alpha\delta}_{d\delta}
       \right),
\end{align}
where repeated indices in each term are summed over.

The VEV of the 75-dimensional scalar representation is
\begin{align}
\langle s\rangle
=v_{75}.
\end{align}
It is this VEV that breaks $SU(5)$ directly down to $SU(3) \times SU(2) \times SU(1)$. Due to this symmetry breaking $X$ and $Y$ vector bosons of $SU(5)$ acquire superheavy mass $m_{X,Y}$. The exact relation between $m_{X,Y}$ and $v_{75§}$ reads
\begin{align}
  m^2_{X,Y}=  48 g^2_\mathrm{GUT}  v_{75}^2,
\end{align}
where $g_\mathrm{GUT}$ is a gauge coupling constant of $SU(5)$ at the unification scale. 

The breaking of $SU(3) \times SU(2) \times U(1)$ down to $SU(3) \times U(1)_\mathrm{em}$ is subsequently accomplished via a single 5-dimensional scalar representation $5^i$ with the following decomposition
\begin{equation}
5=\Lambda=\Lambda_2(1,2,1/2) + \Lambda_3(3,1,-1/3).
\label{eq:5Higgs}
\end{equation}

The operators that only feature renormalizable contractions of the 75-dimensional scalar representation $75^{ij}_{lk}\equiv \Phi^{ij}_{lk}$ with itself are
\begin{align}
I_2 = \Phi^{ij}_{kl}\Phi^{kl}_{ij}, \qquad
I_3 = \Phi^{ij}_{ef}\Phi^{cd}_{ij}\Phi^{ef}_{cd}, \qquad
I_{4,1} = \Phi^{ab}_{gh}\Phi^{cd}_{ab}
          \Phi^{ef}_{cd}\Phi^{gh}_{ef}, \qquad
I_{4,2} = \Phi^{ab}_{cd}\Phi^{cd}_{af}
          \Phi^{ef}_{gh}\Phi^{gh}_{eb},
\end{align}
so that the full potential of the $75$-dimensional scalar representation reads
\begin{equation}
V_{75}= -\frac{\mu^2}{2}I_2
+\frac{\lambda_0}{4}I_2^2
+\frac{\lambda_1}{2}I_{4,1}
+\frac{\lambda_2}{2}I_{4,2}
+\frac{\mu^\prime}{3}I_3. \label{eq:potential}
\end{equation}
The computed squares of the masses of the scalars in $\Phi^{ij}_{lk}$ are accordingly given in Table \ref{tab:masses}.
\begin{longtable}{@{}cccl@{}}
\toprule
field & Sub-multiplet & multiplicity & mass squared \\
\midrule
\endhead
$\Phi_1$ & \((1,1,0)\) & 1 & \(\frac{16}{3}v_{75}\left[\mu^\prime
       +v_{75}(54\lambda_0+30\lambda_1+28\lambda_2)\right]\) \\
$\Phi_{8,1}$ & \((8,1,0)\) & 8 & \(-\frac{4}{3}v_{75}\left[\mu^\prime
       +4v_{75}(3\lambda_1+\lambda_2)\right]\) \\
$\Phi_{8,3}$ & \((8,3,0)\) & 24 & \(-\frac{4}{3}v_{75}\left[5\mu^\prime
       +4v_{75}(3\lambda_1+\lambda_2)\right]\) \\
$\Phi_{3,2}+\Phi_{\overline{3},2}$ & \((3,2,-5/6)+\mathrm{c.c.}\) & 12 & \(0\) \\
$\Phi_{6,2}+\Phi_{\overline{6},2}$ & \((\overline{6},2,-5/6)+\mathrm{c.c.}\) & 24 & \(-\frac{8}{3}v_{75}\left[\mu^\prime
       +v_{75}(12\lambda_1+5\lambda_2)\right]\) \\
$\Phi_{\overline{3},1} + \Phi_{3,1}$ & \((\overline{3},1,-5/3)+\mathrm{c.c.}\) & 6 & \(\frac{16}{3}v_{75}\left[\mu^\prime
       +v_{75}(12\lambda_1-\lambda_2)\right]\) \\
\bottomrule
\caption{Mass squared spectrum of the SM multiplets in $\Phi^{ij}_{lk}$.} \label{tab:masses} 
\end{longtable}

The multiplets  $\Phi_{3,2}$ and $\Phi_{\overline{3},2}$, as shown in Table \ref{tab:masses},  acquire zero mass after one imposes a stationary condition. These are the Goldstone bosons that provide necessary degrees of freedom in order for $X$ and $Y$ gauge bosons to become massive. Our derivation  also reveals that Ref.\ \cite{Hubsch:1984qi} has a wrong overall sign for the mass squared of $\Phi_{8,1}$ multiplet.
Table \ref{tab:masses} allows one to infer existence of a single mass squared relation that reads
\begin{equation}
\label{eq:massrelation_a}
  24m^2_{\Phi_{8,1}} + m^2_{\Phi_{\overline 3,1}}
  - 10m^2_{\Phi_{\overline 6,2}}=0.
\end{equation}
This relation is  relevant for a study of gauge coupling unification that we turn our attention to next. 

Note that if $75$-dimensional representation is assigned a $Z_2$-odd charge, under which $75\to -75$, the cubic term $I_3$ in Eq.~\eqref{eq:potential} would be absent. This scenario is equivalent to simply setting $\mu^\prime=0$ in Table~\ref{tab:masses}, yielding a specific limit where $m_{\Phi_{8,1}}=m_{\Phi_{8,3}}$.

%%%%%%%%%%%%%%%%%%%%%%%%%%%%%%%%%%%%%%%%%%%%%%%%
\subsection{Gauge coupling unification with $75$}
We are finally in a position to analyze gauge coupling unification of this setup. What we are interested in is the highest possible unification scale that one can have. This is particularly useful because the mass $m_{X,Y}$ of the gauge fields $X$ and $Y$ that mediate proton decay in $SU(5)$ can be safely identified with the scale of gauge coupling unification that we simply denote with $m_\mathrm{GUT}$. So, if the theory is to be viable, $m_\mathrm{GUT}$ needs to be high enough in order not to generate rapid decay of nucleons. But, in order for the theory to be testable though nucleon decay signatures, at least in principle, the scale of unification should not be exceedingly large. 

To find the highest possible unification scale $m_{\mathrm{GUT}}$ we explicitly assume that the SM gauge couplings $g_1$, $g_2$, and $g_3$, associated with $U(1)$, $SU(2)$, and $SU(3)$, respectively, unify, at the one-loop level, into a single gauge coupling $g_\mathrm{GUT}$ at $m_{\mathrm{GUT}}$. This allows us to replace three renormalization group equations for $g_1$, $g_2$, and $g_3$ with two equations that solely depend on experimentally measured values of the gauge coupling constants at the $Z$ boson mass scale $m_Z$ and the mass spectrum of all the fields of the theory that reside between $m_Z$ and $m_\mathrm{GUT}$. These equations  are~\cite{Giveon:1991zm}   
\begin{align}
\label{eq:x}
\frac{B_{23}}{B_{12}}&=\frac{5}{8}
\frac{\sin^2
\theta_W-\alpha(m_Z)/\alpha_S(m_Z)}{3/8-\sin^2 \theta_W}=0.71745\,,\\
\label{eq:y}
\ln \frac{m_{\mathrm{GUT}}}{m_Z}&=\frac{16 \pi}{5
\alpha(m_Z)} \frac{3/8-\sin^2 \theta_W}{B_{12}}=\frac{184.828}{B_{12}}\,,
\end{align}
where coefficients $B_{ij}$ are defined via $B_{ij}=\sum_{J} (b^J_{i}-b^J_{j}) r_{J}=\sum_{J} b^J_{ij} r_{J}$. Here,  $b^J_{i}$ are the $\beta$-function coefficients of a particle $J$ with mass $m_J$ and $r_J=\ln(m_{\mathrm{GUT}}/m_{J})/ \ln (m_{\mathrm{GUT}}/m_{Z})$. Note that $J$ goes through all  the fermions and the gauge bosons of the SM as well as all the  scalars in $75$- and $5$-dimensional representations. These scalar multiplets are $\Lambda_2$, $\Lambda_3$, $\Phi_{8,1}$, $\Phi_{8,3}$, $\Phi_{6,2}$, $\Phi_{\overline{6},2}$, $\Phi_{\overline{3},1}$, and  $\Phi_{3,1}$. We list $\beta$-function coefficients $b^J_i$~\cite{Gross:1973id,Politzer:1973fx} and the associated values for $b^J_{12}$ and $b^J_{23}$ for all the scalars and the fermions of the model in unshaded part of Table~\ref{tab:ParticleContent}. 

To generate numbers presented in Eqs.\ \eqref{eq:x} and \eqref{eq:y} we use the following input at  $m_Z(=91.1888$\,GeV~\cite{ParticleDataGroup:2024cfk}) scale~\cite{Antusch:2025fpm}:
\begin{align}
    g_1 = 0.461228 \pm 0.000026 ,\quad g_2 = 0.65096 \pm 0.00004 ,\quad g_3 = 1.2123 \pm 0.0046 .
\end{align}
Since the experimental errors are negligible for our purpose, we only use the central values of $g_i$s to obtain
$\left(\sin^2\theta_W(m_Z), \alpha^{-1}(m_Z), \alpha^{-1}_S(m_Z) \right) =  \left(   0.231486, 128.108, 8.55046\right)$, where $\alpha_S=g_3^2/(4 \pi)$ is a strong coupling while $\alpha$ is an electromagnetic coupling constant.

%%%%%%%%%%%%%%%%%%
\begin{figure}[t!]
\centering
\includegraphics[width=0.8\textwidth]{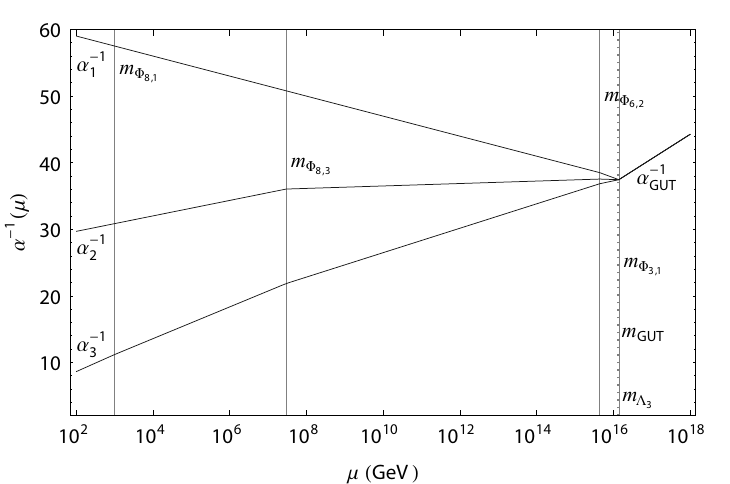}
\caption{Mass spectrum that yields gauge coupling unification for the maximal possible $m_\mathrm{GUT}$ in a scenario with 75-dimensional and 5-dimensional  scalar representations. The fermion sector comprises  $10^A_F$ and $ \overline{5}^A_F$, where $A=1,2,3$.} \label{fig:75_a}
\end{figure}
%%%%%%%%%%%%%%%%%%  

We note that the sign of the contribution of any field $J$ towards $B_{ij}$s is purely determined by the sign of its coefficients $b^J_{ij}$s. To maximize $m_\mathrm{GUT}$ one would thus need a field $J$ with negative $b^J_{12}$ contribution, as evident from a denominator of Eq.\ \eqref{eq:y}. Table~\ref{tab:ParticleContent} shows that there is only one such scalar multiplet beside the Higgs doublet $\Lambda_2$ and that multiplet is $\Phi_{8,3}$. Since Eqs.\ \eqref{eq:x} and \eqref{eq:y} need to be solved simultaneously, one would ideally like to have a multiplet $J$ that has positive $b^J_{23}$ and negative $b^J_{12}$ as the ratio $B^\mathrm{SM}_{23}/B^\mathrm{SM}_{12} \approx 0.5$ for the SM significantly undershoots experimental value presented on the right hand side of Eq.\ \eqref{eq:x}. Again, the only field that has both of these properties is $\Phi_{8,3}$. In fact, $\Phi_{8,3}$ is so efficient in satisfying Eq.\ \eqref{eq:x} that it overshoots experimentally required value for $B_{23}/B_{12}(=0.71745)$ long before it can reach $m_Z$ scale. In other words, $\Phi_{8,3}$  cannot significantly affect $m_\mathrm{GUT}$ through Eq.\ \eqref{eq:y} on its own as it can fully satisfy Eq.\ \eqref{eq:x} even when its mass is relatively close to $m_\mathrm{GUT}$. This is where multiplet $\Phi_{8,1}$ steps in. Namely, $\Phi_{8,1}$ cannot directly affect $m_\mathrm{GUT}$ as it has $b_{12}=0$, but it does that indirectly as it allows multiplet $\Phi_{8,3}$ to go to even lower scales with respect to $m_\mathrm{GUT}$ since it can compensate positivity of the $\Phi_{8,3}$ contribution towards $B_{23}$. In short, there are only two scalars that are relevant for the maximization of the unification scale $m_\mathrm{GUT}$ or, equivalently, minimization of $B_{12}$ coefficient of Eq.\ \eqref{eq:y} in the scenario with 75-dimensional scalar representation. These scalar multiplets are $\Phi_{8,1}$ and $\Phi_{8,3}$, where the former prefers to be as close as possible to the electroweak scale. 

Our discussion is supported by the one-loop level numerical analysis of the gauge coupling unification. If we freely vary all the relevant masses of the fields in 75-dimensional representation between $1$\,TeV and $m_\mathrm{GUT}$ while taking into account Eq.\ \eqref{eq:massrelation_a}, we find that the  unification scale is limited to the following range: $1.6 \times 10^{15}\,\mathrm{GeV} \leq m_\mathrm{GUT} \leq 1.3 \times 10^{16}\,\mathrm{GeV}$. We accordingly present in Fig.\ \ref{fig:75_a} unification scenario that corresponds to the maximal possible unification scale of $m_\mathrm{GUT} = 1.3 \times 10^{16}\,\mathrm{GeV}$. 

Note that $\Lambda_2 \in 5$ is always at the electroweak scale as it corresponds to the Higgs doublet of the SM, whereas Fig.\ \ref{fig:75_a} places $\Lambda_3 \in 5$ at $m_\mathrm{GUT}$. This is the source of what is known as the doublet-triplet splitting problem~\cite{Randall:1995sh,Yamashita:2011an}. The model with 75-dimensional representation certainly has enough parameters to accomplish this tremendous mass splitting through fine-tuning as it has two linearly independent operators $\Lambda_{i}^{*}\Lambda^{i}\Phi_{lm}^{jk}\Phi_{jk}^{lm}$ and $ \Lambda_{j}^{*}\Lambda^{i}\Phi_{lm}^{jk}\Phi_{ki}^{lm}$ that could be canceled against $m^2 \Lambda_{i}^{*}\Lambda^{i}$ contraction to yield light doublet $\Lambda_2$ and superheavy triplet $\Lambda_3$. It is interesting to note, though, that there is another operator basis, where one of the contractions between $75$-dimensional scalar representation and $5$-dimensional scalar representation yields mass solely for the triplet component $\Lambda_3 \in 5$. This operator  reads
\begin{equation}
    \mathcal{O} = \lambda \epsilon_{ijklm}\Lambda_{n}^{*}\Phi_{po}^{kl}\Phi_{qr}^{mo}\Lambda^{j}\epsilon^{inpqr},
    \label{eq:Triplet_only}
\end{equation}
whereas the square of the triplet mass comes out to be
$m^2_{\Lambda_3}=96 \lambda v_{75}^2 
$. We atribute this curiosity to the following identity: 
\begin{equation}
     \epsilon_{ijklm}\Lambda_{n}^{*}\Phi_{po}^{kl}\Phi_{qr}^{mo}\Lambda^{j}\epsilon^{inpqr} = 2 \Lambda_{i}^{*}\Lambda^{i}\Phi_{lm}^{jk}\Phi_{jk}^{lm} + 6 \Lambda_{j}^{*}\Lambda^{i}\Phi_{lm}^{jk}\Phi_{ki}^{lm}. 
\end{equation}

As one can see from Fig.\ \ref{fig:75_a}, only  $\Phi_{8,1}$ and $\Phi_{8,3}$ reside below the unification scale, in accordance with our qualitative discussion. We furthermore emphasis that any departure from the exact mass spectrum of Fig.\ \ref{fig:75_a} can only yield lower value of $m_\mathrm{GUT}$. Our analysis of $m_\mathrm{GUT}$ should thus be considered as the most conservative one.

\begin{figure}[t!]
\centering
\includegraphics[width=1\textwidth]{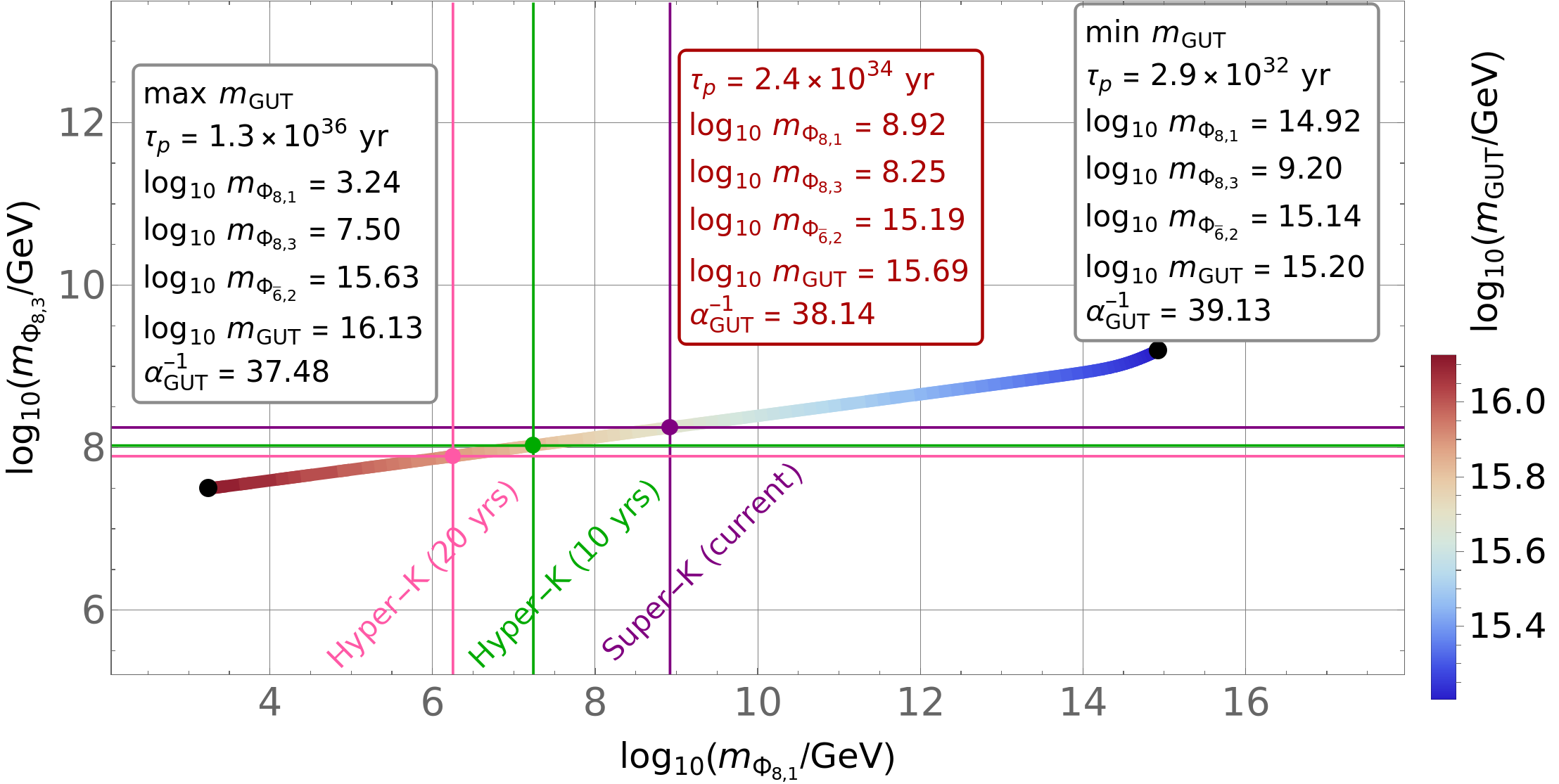}
\caption{Proton decay estimation in a scenario with 75-dimensional and 5-dimensional  scalar representations. Relevant partial lifetime bounds are summarized in Table~\ref{tab:nucleon_decay}. } \label{fig:moneyplot}
\end{figure}

The model clearly predicts that $m_\mathrm{GUT}$ must reside within a very narrow range, where part of the available parameter space is already in conflict with matter stability.
To illustrate this point, we present in Fig.\ \ref{fig:moneyplot} our estimation of the proton decay lifetime mediated by $X$ and $Y$ gauge bosons. Again,  we identify the superheavy gauge boson mass $m_{X,Y}$ with the unification scale $m_\mathrm{GUT}$. Hence, the proton lifetime estimation~\cite{Langacker:1980js} reads  
\begin{align}
\tau_p\sim
\frac{16\pi^2  m^4_{X,Y}}{g^4_{\textrm{GUT}}  m^5_p},
\end{align}
where $m_p$ is a proton mass. The most stringent constraint on partial proton decay lifetime arises from the decay mode $p\rightarrow \pi^0\,e^+$ and the associated current experimental bound and future sensitivities are summarized in Table \ref{tab:nucleon_decay}.

%%%%%%%%%%%%%%%%%%
%%%%%%%%%%%%%%%%%% 
 
\begin{table}[th!]
\centering
\begin{tabular}{|c|c|c|c|}\hline
Decay channel  & Current bound $\tau_p$ [yrs] & 10 yr sensitivity $\tau_p$ [yrs]  & 20 yr sensitivity $\tau_p$ [yrs] 
\\\hline\hline
$p\rightarrow \pi^0\,e^+$  &  $2.4\times 10^{34}$ \cite{Super-Kamiokande:2020wjk} &  $7.8\times 10^{34}$ \cite{Hyper-Kamiokande:2018ofw} &  $1.56\times 10^{35}$ \cite{Dev:2022jbf}   \\ \hline
\end{tabular}
\caption{ Current bound and future expectations for $p\rightarrow \pi^0\,e^+$ decay channel. }\label{tab:nucleon_decay}
\end{table}

It is clear from  
Fig.\ \ref{fig:moneyplot} that Super-K yields the following upper limits on $m_{\Phi_{8,1}}$ and $m_{\Phi_{8,3}}$:  
\begin{align}
\textrm{Super-K}:\quad &  m_{\Phi_{8,1}} \lesssim  8.3\times 10^8\,\mathrm{GeV} \quad \mathrm{\&} \quad   m_{\Phi_{8,3}} \lesssim  1.7\times 10^8\,\mathrm{GeV}    
\end{align}
Fig.\ \ref{fig:moneyplot} also confirms that $m_{\Phi_{8,3}}$ is constrained to be in a very narrow range since $\Phi_{8,3}$ efficiently satisfies Eq.\ \eqref{eq:x} and thus cannot reside below $10^{7.5}$\,GeV under any circumstances, whereas $\Phi_{8,1}$ prefers to be in the TeV region in order to maximize $m_\mathrm{GUT}$. Note that if one implements $Z_2$-odd symmetry $75 \to -75$ that implies $m_{\Phi_{8,1}}=m_{\Phi_{8,3}}$, only a small region of the full parameter space of Fig.\ \ref{fig:moneyplot} that is near $m_{\Phi_{8,1}}=m_{\Phi_{8,3}} \simeq 10^8$\,GeV would be viable for gauge coupling unification. Intriguingly, this $Z_2$-odd window would then be fully tested by Hyper-K.

One might object that Fig.\ \ref{fig:moneyplot} is not overly illuminating since the scenario where 24-dimensional scalar representation is simply replaced with 75-dimensional representation cannot break mass degeneracy between the down-type quarks and the charged leptons. This means that the fermion mass spectrum is not realistic and it is thus impossible to  properly determine all unitary transformations that take the SM fermions from the flavor basis into the mass eigenstate basis. All this is true but our point is that the theory with 75-dimensional scalar representation can certainly provide viable gauge coupling unification with relatively accessible $m_\mathrm{GUT}$ in contrast to the Georgi-Glashow model that cannot do even that. However, as we are also interested in a fully viable model we provide in what follows an analysis of one such scenario. 

%%%%%%%%%%%%%%%%%%%%%%%%%%%%%
\section{$75$ with vectorlike $10_F+\overline{10}_F$}
%%%%%%%%%%%%%%%%%%%%%%%%%%%%%
\label{sec:two} 
To generate realistic fermion masses in $SU(5)$ one can introduce additional scalar representations~\cite{Georgi:1979df}, vectorlike fermions, or both. We opt to investigate vectorlike fermion extensions in what follows since these  introduce fewer parameters when compared with extensions that solely rely on scalars. 

In the theory with 24-dimensional scalar representation one can add three different types of vectorlike fermions to break the Georgi-Glashow model prediction for the mass degeneracy between the down-type quarks and the charged leptons. These fermion pairs comprise $5_F+\overline{5}_F$~\cite{Babu:2012pb,Dorsner:2014wva,Antusch:2023mqe}, $10_F+\overline{10}_F$~\cite{Calibbi:2022wko,Antusch:2023mqe}, and $15_F+\overline{15}_F$~\cite{Dorsner:2019vgf,Dorsner:2021qwg,Antusch:2023jok,Dorsner:2024jiy}. Of course, one can introduce numerous copies of these vectorlike fields and/or combinations of different types but a single vectorlike addition is already sufficient to produce viable masses of the SM charged fermions.

The situation is different in the theory with 75-dimensional scalar representation since there is only one viable option if one is to introduce vectorlike fermions due to antisymmetric tensorial nature of $75^{ij}_{kl}$ and the fact that $75^{ij}_{jk}=0$. Namely, one can only extend fermion sector with a vectorlike $10_F+\overline{10}_F$ pair. All other options are simply ruled out on the group theoretical grounds. This makes 75-dimensional extension rather unique.  

This extension is even more exceptional in the sense that if 75-dimensional representation is replaced with 200-dimensional representation, one cannot  correct the bad mass relations using vectorlike fermions at all. Namely, even though one would expect that an inclusion of a  vectorlike $15_F+\overline{15}_F$ pair might work in the scenario with 200-dimensional scalar representation, it is easy to show that it is impossible to introduce necessary mixing of vectorlike addition with the SM fermions. This is due to the fact that there is no equivalent contraction to $10^{ij}_F \overline{15}_{Fik} 24^k_j$ that one could write down in the scenario with $200^{ij}_{kl}$ due to its tensorial properties.

We proceed to explicitly show how one breaks mass degeneracy between the down-type quarks and the charged leptons. We, for definiteness, list in Table~\ref{tab:ParticleContent} both the scalar and the fermion  content of the scenario under consideration. The fermion sector of the model thus  comprises  $10^A_F$, $ \overline{5}^a_F$, and $\overline{10}_F$, where $A(=1,2,3,4)$ and $a(=1,2,3)$ are flavor indices. The scalar sector consists of one 75-dimensional and one 5-dimensional representation.

The lagrangian of the Yukawa sector of the scenario is
\begin{align}
&\mathcal{L}_Y\supset  Y_{10}^{AB}10_F^A 10^B_F5+Y_{5}^{Aa} 10^A_F  \overline 5^a_F 5^*   + y \overline{10}_F\overline{10}_F5^* + \left(m_A  + \zeta_A 75\right)   10_F^A  \overline{10}_F,\label{eq:Yukawa}
\end{align}
where $A,B=1,2,3,4$ and $a=1,2,3$. 
Note that the SM multiplets of the model are embedded into $SU(5)$ representations as follows 
\begin{align*}
\overline{5}_{F}^a=\begin{pmatrix}
d^c_r&d^c_b&d^c_g&e& -\nu
\end{pmatrix}_a^T,
\end{align*}
\begin{align*}
10^A_F=\frac{1}{\sqrt{2}} \begin{pmatrix}
0&u^c_g&-u^c_b&u_r&d_r\\
-u^c_g&0&u^c_r&u_b&d_b\\
u^c_b&-u^c_r&0&u_g&d_g\\
-u_r&-u_b&-u_g&0&e^c\\
-d_r&-d_b&-d_g&-e^c&0
\end{pmatrix}_A,\;\;\;
\overline{10}_F=\frac{1}{\sqrt{2}} \begin{pmatrix}
0&U_g&-U_b&U^c_r&D^c_r\\
-U_g&0&U_r&U^c_b&D^c_b\\
U_b&-U_r&0&U^c_g&D^c_g\\
-U^c_r&-U^c_b&-U^c_g&0&E\\
-D^c_r&-D^c_b&-D^c_g&-E&0
\end{pmatrix}.
\end{align*}
We will, from here on, always suppress the color subscripts $g$, $r$, and $b$.

Since $     \overline{10}_{F}= Q^c(\overline{3},2,-1/6) + U(3,1,2/3) + E(1,1,-1)$, with $Q^c= (U^c, D^c)^T$, in accordance with Table \ref{tab:ParticleContent},
we find that the last two terms in Eq.\ \eqref{eq:Yukawa} yield the following masses for the vectorlike fermions:
\begin{align}
&m_{U} = m_4  + 2\zeta_4 v_{75},\quad m_{Q^c} = m_4  - 2\zeta_4 v_{75},\quad m_{E} = m_4  + 6\zeta_4 v_{75}.
\end{align}
These lead to a mass relation
\begin{align}
&m_{U} = \frac{{m_{Q^c}} + m_{E}}{2},
\label{eq:massVLfermions}    
\end{align}
where the fermion masses should be treated as complex parameters.
If one uses  $24$-dimensional scalar representation instead of $75$-dimensional representation, one obtains Eq.\ \eqref{eq:massVLfermions}
but with $U \leftrightarrow Q^c$~\cite{Antusch:2023mqe}.

If we take the VEV of a 5-dimensional scalar representation to be
\begin{align} 
\langle 5\rangle=\frac{v}{\sqrt{2}},
\end{align}
we obtain  mass matrices $\mathcal{M}_D$, $\mathcal{M}_E$, and  $\mathcal{M}_U$ for the fermions of the theory via
\begin{align}
\mathcal{L}_Y\supset & 
\begin{pmatrix}
d_1&d_2&d_3&d_4    
\end{pmatrix}
\mathcal{M}_D
\begin{pmatrix}
d_1^c&d_2^c&d_3^c&D^c     
\end{pmatrix}^T
+\begin{pmatrix}
e_1&e_2&e_3&E    
\end{pmatrix}
\mathcal{M}_E
\begin{pmatrix}
e_1^c&e_2^c&e_3^c&e_4^c     
\end{pmatrix}^T
\nonumber \\
&+\begin{pmatrix}
u_1&u_2&u_3&u_4&U    
\end{pmatrix}
\mathcal{M}_U
\begin{pmatrix}
u_1^c&u_2^c&u_3^c&u_4^c&U^c     
\end{pmatrix}^T\;,
\end{align}
where these matrices are explicitly given as follows:
\begin{align}
\label{eq:ddd}
&\mathcal{M}_D=
\begin{pmatrix}
\underbrace{\left(Y_{5}\right)^{ab}\frac{v}{2}}_{3\times 3}
&
\underbrace{m_a-\zeta_a v_{75}}_{3\times 1}  
\\
\underbrace{\left(Y_{5}\right)^{4b}\frac{v}{2}}_{1\times 3}
&
\underbrace{m_4-\zeta_4 v_{75}}_{1\times 1} 
\end{pmatrix},
\\
\label{eq:eee}
&\mathcal{M}_E=
\begin{pmatrix}
\underbrace{\left(Y_{5}^T\right)^{ab}\frac{v}{2}}_{3\times 3}
&
\underbrace{\left(Y_{5}^T\right)^{a4}\frac{v}{2}}_{3\times 1}
\\
\underbrace{m_b+ 3\zeta_b v_{75}}_{1\times 3}  
&
\underbrace{m_4+3\zeta_4 v_{75}}_{1\times 1} 
\end{pmatrix},
\\
&\mathcal{M}_U=
\begin{pmatrix}
\underbrace{\sqrt{2}\left(Y_{10}+Y_{10}^T\right)^{AB}v}_{4\times 4}
&
\underbrace{m_A-\zeta_A v_{75}}_{4\times 1}
\\
\underbrace{m_B+\zeta_B v_{75}}_{1\times 4}  
&
\underbrace{2\sqrt{2}y v}_{1\times 1} 
\end{pmatrix}.
\end{align}
Again, $A,B=1,2,3,4$ and $a,b=1,2,3$. It is clear from Eqs.\ \eqref{eq:ddd} and \eqref{eq:eee} that the unwanted mass degeneracy between the down-type quarks and the charged leptons can easily be broken.

\textbf{Neutrino mass}:  
Since we are interested in a scenario that can accommodate all the masses of the SM fermions we opt for implementation of the Type I seesaw mechanism~\cite{Minkowski:1977sc,Gell-Mann:1979vob,Yanagida:1979as,Schechter:1980gr,Glashow:1979nm,Mohapatra:1979ia} through introduction of at least two generations of right-handed neutrinos that are singlets of $SU(5)$. One particular motivation for employing the Type I seesaw is that it has no effects on the gauge coupling unification whatsoever.

%%%%%%%%%%%%%%%%%%%%%%%%%%%%%%%%%%%%%%%%%%%%%%%%
\subsection{Gauge coupling unification with $75$ and  $10_F+\overline{10}_F$}
\begin{figure}[th!]
\centering
\includegraphics[width=0.8\textwidth]{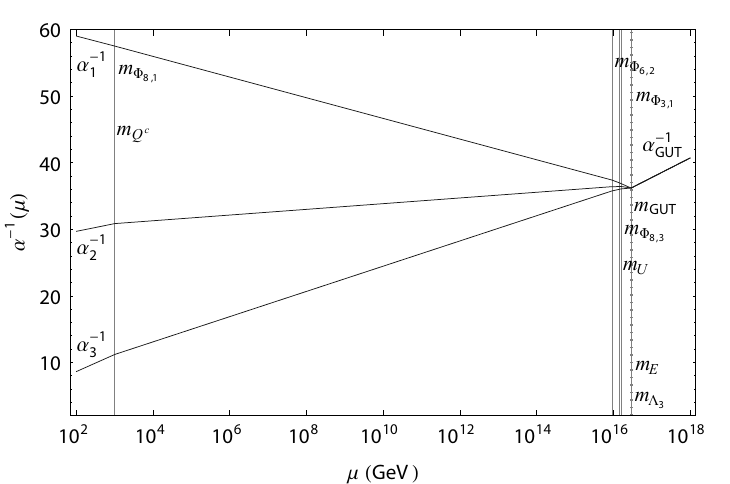}
\caption{Mass spectrum that yields gauge coupling unification for the maximal possible $m_\mathrm{GUT}$ in a scenario with 75-dimensional and 5-dimensional  scalar representations. The fermion sector  comprises  $10^A_F$, $ \overline{5}^a_F$, and $\overline{10}_F$, where $A=1,2,3,4$ and $a=1,2,3$.} \label{fig:75_b}
\end{figure}

We are, once again, interested in finding the largest possible scale of unification. The shaded content of Table \ref{tab:ParticleContent} reveals that the fermions that reside in a vectorlike pair that transform under $SU(3)\times SU(2) \times U(1)$ as the SM quark doublets have negative $b_{12}$ and positive $b_{23}$ coefficients. They are thus ideal candidates to satisfy Eq.\ \eqref{eq:x} and, in the process, maximize $m_\mathrm{GUT}$ through Eq.\ \eqref{eq:y}. In fact, numerical analysis prefers contribution of $Q^c$   over the contribution of scalar multiplet $\Phi_{8,3}$ that is simply too efficient in satisfying Eq.\ \eqref{eq:x}. The scalar multiplet $\Phi_{8,1}$ again helps indirectly to maximize $m_\mathrm{GUT}$ as it allows vectorlike fermions to go as low as possible towards $1$\,TeV scale. Our study shows that the maximal possible unification scale is given by the limit $m_\mathrm{GUT} \leq 2.9 \times 10^{16}$\,GeV, where both scalar multiplet $\Phi_{8,1}$ and vectorlike fermion multiplet $Q^c$ are at 1\,TeV scale. We accordingly present in Fig.\ \ref{fig:75_b} the mass spectrum that corresponds to the maximal possible $m_\mathrm{GUT}$ within this scenario. Any departure from this particular mass spectrum would yield lower unification scale.

It is interesting to note that the Georgi-Glashow model extended with one pair of vectorlike fermions in $10_F+\overline{10}_F$ or $15_F+\overline{15}_F$ yields maximal unification scale that is a factor of two larger~\cite{Antusch:2023mqe} than the present scenario with the $75$-dimensional representation and a vectorlike $10_F+\overline{10}_F$ pair.
%%%%%%%%%%%%%%%%%%
%%%%%%%%%%%%%%%%%% 

%%%%%%%%%%%%%%%%%%%%%%%%%%%%%
\section{Conclusions}
%%%%%%%%%%%%%%%%%%%%%%%%%%%%%
\label{sec:three}
We study phenomenology of two specific modifications of the  Georgi-Glashow $SU(5)$ setup in this manuscript.
The first modification corresponds to a  renormalizable $SU(5)$ grand unified scenario in which the  24-dimensional scalar representation of the original Georgi-Glashow model is simply replaced by a 75-dimensional scalar representation. We show that, despite the quantitative enlargement of the symmetry breaking sector, the scenario turns out to be highly predictive.  We demonstrate that this scenario achieves gauge coupling unification, unlike the minimal 24-dimensional Georgi-Glashow model, and we derive the range of allowed unification scales to be  $1.6\times10^{15}\,\mathrm{GeV} \lesssim m_{\rm GUT} \lesssim 1.3\times10^{16}\,\mathrm{GeV}$. We furthermore demonstrate  that there are only two scalar multiplets that are relevant for viable gauge coupling unification. Both of these multiplets are part of the 75-dimensional representation. One is an octet of $SU(3)$ and a triplet of $SU(2)$, whereas the other multiplet simply transforms as an octet of $SU(3)$. We furthermore confront the mass spectrum of this scenario with the current and the projected Super-Kamiokande/Hyper-Kamiokande sensitivities to $p\to\pi^0 e^+$ to establish explicit bounds on these two multiplets. Our findings show that the part of the parameter space of this scenario is already excluded by the current nucleon-decay search results. 

Since a simple 75-dimensional scalar extension inherits unrealistic mass degeneracy between the down-type quarks and the charged leptons, we extend it with a single vectorlike $10_F+\overline{10}_F$ pair. It turns out that the $10_F+\overline{10}_F$ vectorlike addition is unique due to antisymmetric tensorial structure of $75$-dimensional representation. We proceed to show that this extension can simultaneously generate viable charged-fermion masses and gauge coupling unification, where the maximal unification scale is given by $m_{\rm GUT}\le 2.9\times10^{16}$\,GeV. The proposed framework can thus serve as a  phenomenologically viable alternative to the standard Georgi-Glashow paradigm. It is a rather unique scenario and a large part of its parameter space is accessible to current and future experiments on nucleon stability.

%%%%%%%%%%%%%%%%%%%%%%%%%%%%%
\section*{Acknowledgments}
%%%%%%%%%%%%%%%%%%%%%%%%%%%%%
I.D.\ acknowledges the financial support
from the project ProPuBFO-1.1.3.2026.  S.S.\  acknowledges the financial support from the Slovenian Research Agency (research core funding No.\ P1-0035). 

\bibliographystyle{style}
\bibliography{references}

\providecommand{\href}[2]{#2}\begingroup\raggedright\begin{thebibliography}{10}

\bibitem{Georgi:1974sy}
H.~Georgi and S.~L. Glashow, ``{Unity of All Elementary Particle Forces},''
\href{http://dx.doi.org/10.1103/PhysRevLett.32.438}{{\em Phys. Rev. Lett.} {\bfseries 32} (1974) 438--441}.
%%CITATION = PRLTA,32,438;%%.

\bibitem{Hubsch:1984pg}
T.~Hubsch and S.~Pallua, ``{Symmetry Breaking Mechanism in an Alternative SU(5) Model},'' \href{http://dx.doi.org/10.1016/0370-2693(84)91659-9}{{\em Phys. Lett. B} {\bfseries 138} (1984) 279--282}.

\bibitem{Hubsch:1984qi}
T.~Hubsch, S.~Meljanac, and S.~Pallua, ``{A Nonminimal SU(5) Model},'' \href{http://dx.doi.org/10.1103/PhysRevD.31.2958}{{\em Phys. Rev. D} {\bfseries 31} (1985) 2958}.

\bibitem{Hubsch:1984zi}
T.~Hubsch, S.~Meljanac, and S.~Pallua, ``{Symmetry Breaking of SU($n$) Gauge Theories to Maximal Regular Subgroups and Fourth Rank Tensors},'' \href{http://dx.doi.org/10.1103/PhysRevD.31.352}{{\em Phys. Rev. D} {\bfseries 31} (1985) 352}.

\bibitem{Giveon:1991zm}
A.~Giveon, L.~J. Hall, and U.~Sarid, ``{SU (5) unification revisited},'' \href{http://dx.doi.org/10.1016/0370-2693(91)91289-8}{{\em Phys. Lett. B} {\bfseries 271} (1991) 138--144}.

\bibitem{Gross:1973id}
D.~J. Gross and F.~Wilczek, ``{Ultraviolet Behavior of Non-Abelian Gauge Theories},'' \href{http://dx.doi.org/10.1103/PhysRevLett.30.1343}{{\em Phys. Rev. Lett.} {\bfseries 30} (1973) 1343--1346}.

\bibitem{Politzer:1973fx}
H.~D. Politzer, ``{Reliable Perturbative Results for Strong Interactions?},'' \href{http://dx.doi.org/10.1103/PhysRevLett.30.1346}{{\em Phys. Rev. Lett.} {\bfseries 30} (1973) 1346--1349}.

\bibitem{ParticleDataGroup:2024cfk}
{\bfseries Particle Data Group} Collaboration, S.~Navas {\em et~al.}, ``{Review of particle physics},'' \href{http://dx.doi.org/10.1103/PhysRevD.110.030001}{{\em Phys. Rev. D} {\bfseries 110} no.~3, (2024) 030001}.

\bibitem{Antusch:2025fpm}
S.~Antusch, K.~Hinze, and S.~Saad, ``{Updated running quark and lepton parameters at various scales},'' \href{http://dx.doi.org/10.1103/fdcc-ycph}{{\em Phys. Rev. D} {\bfseries 113} no.~9, (2026) 095011}, \href{http://arxiv.org/abs/2510.01312}{{\ttfamily arXiv:2510.01312 [hep-ph]}}.

\bibitem{Randall:1995sh}
L.~Randall and C.~Csaki, ``{The Doublet - triplet splitting problem and Higgses as pseudoGoldstone bosons},'' in {\em {International Workshop on Supersymmetry and Unification of Fundamental Interactions (SUSY 95)}}, pp.~99--109.
\newblock 3, 1995.
\newblock \href{http://arxiv.org/abs/hep-ph/9508208}{{\ttfamily arXiv:hep-ph/9508208}}.

\bibitem{Yamashita:2011an}
T.~Yamashita, ``{Doublet-Triplet Splitting in an SU(5) Grand Unification},'' \href{http://dx.doi.org/10.1103/PhysRevD.84.115016}{{\em Phys. Rev. D} {\bfseries 84} (2011) 115016}, \href{http://arxiv.org/abs/1106.3229}{{\ttfamily arXiv:1106.3229 [hep-ph]}}.

\bibitem{Langacker:1980js}
P.~Langacker, ``{Grand Unified Theories and Proton Decay},'' \href{http://dx.doi.org/10.1016/0370-1573(81)90059-4}{{\em Phys. Rept.} {\bfseries 72} (1981) 185}.

\bibitem{Super-Kamiokande:2020wjk}
{\bfseries Super-Kamiokande} Collaboration, A.~Takenaka {\em et~al.}, ``{Search for proton decay via $p\to e^+\pi^0$ and $p\to \mu^+\pi^0$ with an enlarged fiducial volume in Super-Kamiokande I-IV},'' \href{http://dx.doi.org/10.1103/PhysRevD.102.112011}{{\em Phys. Rev. D} {\bfseries 102} no.~11, (2020) 112011}, \href{http://arxiv.org/abs/2010.16098}{{\ttfamily arXiv:2010.16098 [hep-ex]}}.

\bibitem{Hyper-Kamiokande:2018ofw}
{\bfseries Hyper-Kamiokande} Collaboration, K.~Abe {\em et~al.}, ``{Hyper-Kamiokande Design Report},'' \href{http://arxiv.org/abs/1805.04163}{{\ttfamily arXiv:1805.04163 [physics.ins-det]}}.

\bibitem{Dev:2022jbf}
P.~S.~B. Dev {\em et~al.}, ``{Searches for baryon number violation in neutrino experiments: a white paper},'' \href{http://dx.doi.org/10.1088/1361-6471/ad1658}{{\em J. Phys. G} {\bfseries 51} no.~3, (2024) 033001}, \href{http://arxiv.org/abs/2203.08771}{{\ttfamily arXiv:2203.08771 [hep-ex]}}.

\bibitem{Georgi:1979df}
H.~Georgi and C.~Jarlskog, ``{A New Lepton - Quark Mass Relation in a Unified Theory},'' \href{http://dx.doi.org/10.1016/0370-2693(79)90842-6}{{\em Phys. Lett. B} {\bfseries 86} (1979) 297--300}.

\bibitem{Babu:2012pb}
K.~S. Babu, B.~Bajc, and Z.~Tavartkiladze, ``{Realistic Fermion Masses and Nucleon Decay Rates in SUSY SU(5) with Vector-Like Matter},'' \href{http://dx.doi.org/10.1103/PhysRevD.86.075005}{{\em Phys. Rev. D} {\bfseries 86} (2012) 075005}, \href{http://arxiv.org/abs/1207.6388}{{\ttfamily arXiv:1207.6388 [hep-ph]}}.

\bibitem{Dorsner:2014wva}
I.~Dorsner, S.~Fajfer, and I.~Mustac, ``{Light vector-like fermions in a minimal SU(5) setup},'' \href{http://dx.doi.org/10.1103/PhysRevD.89.115004}{{\em Phys. Rev.} {\bfseries D89} no.~11, (2014) 115004},
\href{http://arxiv.org/abs/1401.6870}{{\ttfamily arXiv:1401.6870 [hep-ph]}}.
%%CITATION = ARXIV:1401.6870;%%.

\bibitem{Antusch:2023mqe}
S.~Antusch, K.~Hinze, and S.~Saad, ``{Minimal SU(5) GUTs with vectorlike fermions},'' \href{http://dx.doi.org/10.1103/PhysRevD.108.095010}{{\em Phys. Rev. D} {\bfseries 108} no.~9, (2023) 095010}, \href{http://arxiv.org/abs/2308.08585}{{\ttfamily arXiv:2308.08585 [hep-ph]}}.

\bibitem{Calibbi:2022wko}
L.~Calibbi and X.~Gao, ``{Lepton flavor violation in minimal grand unified type II seesaw models},'' \href{http://dx.doi.org/10.1103/PhysRevD.106.095036}{{\em Phys. Rev. D} {\bfseries 106} no.~9, (2022) 095036}, \href{http://arxiv.org/abs/2206.10682}{{\ttfamily arXiv:2206.10682 [hep-ph]}}.

\bibitem{Dorsner:2019vgf}
I.~Dor\v{s}ner and S.~Saad, ``{Towards Minimal $SU(5)$},'' \href{http://dx.doi.org/10.1103/PhysRevD.101.015009}{{\em Phys. Rev. D} {\bfseries 101} no.~1, (2020) 015009}, \href{http://arxiv.org/abs/1910.09008}{{\ttfamily arXiv:1910.09008 [hep-ph]}}.

\bibitem{Dorsner:2021qwg}
I.~Dor\v{s}ner, E.~D\v{z}aferovi\'c-Ma\v{s}i\'c, and S.~Saad, ``{Parameter space exploration of the minimal SU(5) unification},'' \href{http://dx.doi.org/10.1103/PhysRevD.104.015023}{{\em Phys. Rev. D} {\bfseries 104} no.~1, (2021) 015023}, \href{http://arxiv.org/abs/2105.01678}{{\ttfamily arXiv:2105.01678 [hep-ph]}}.

\bibitem{Antusch:2023jok}
S.~Antusch, I.~Dor\v{s}ner, K.~Hinze, and S.~Saad, ``{Fully testable axion dark matter within a minimal SU(5) GUT},'' \href{http://dx.doi.org/10.1103/PhysRevD.108.015025}{{\em Phys. Rev. D} {\bfseries 108} no.~1, (2023) 015025}, \href{http://arxiv.org/abs/2301.00809}{{\ttfamily arXiv:2301.00809 [hep-ph]}}.

\bibitem{Dorsner:2024jiy}
I.~Dor\v{s}ner, E.~D\v{z}aferovi\'c-Ma\v{s}i\'c, S.~Fajfer, and S.~Saad, ``{Gauge and scalar boson mediated proton decay in a predictive SU(5) GUT model},'' \href{http://dx.doi.org/10.1103/PhysRevD.109.075023}{{\em Phys. Rev. D} {\bfseries 109} no.~7, (2024) 075023}, \href{http://arxiv.org/abs/2401.16907}{{\ttfamily arXiv:2401.16907 [hep-ph]}}.

\bibitem{Minkowski:1977sc}
P.~Minkowski, ``{$\mu \to e\gamma$ at a Rate of One Out of $10^{9}$ Muon Decays?},''
\href{http://dx.doi.org/10.1016/0370-2693(77)90435-X}{{\em Phys. Lett.} {\bfseries 67B} (1977) 421--428}.
%%CITATION = PHLTA,67B,421;%%.

\bibitem{Gell-Mann:1979vob}
M.~Gell-Mann, P.~Ramond, and R.~Slansky, ``{Complex Spinors and Unified Theories},'' {\em Conf. Proc. C} {\bfseries 790927} (1979) 315--321, \href{http://arxiv.org/abs/1306.4669}{{\ttfamily arXiv:1306.4669 [hep-th]}}.

\bibitem{Yanagida:1979as}
T.~Yanagida, ``{Horizontal gauge symmetry and masses of neutrinos},''
{\em Conf. Proc.} {\bfseries C7902131} (1979) 95--99.
%%CITATION = CONFP,C7902131,95;%%.

\bibitem{Schechter:1980gr}
J.~Schechter and J.~W.~F. Valle, ``{Neutrino Masses in SU(2) x U(1) Theories},''
\href{http://dx.doi.org/10.1103/PhysRevD.22.2227}{{\em Phys. Rev.} {\bfseries D22} (1980) 2227}.
%%CITATION = PHRVA,D22,2227;%%.

\bibitem{Glashow:1979nm}
S.~Glashow, ``{The Future of Elementary Particle Physics},'' \href{http://dx.doi.org/10.1007/978-1-4684-7197-7\_15}{{\em NATO Sci. Ser. B} {\bfseries 61} (1980) 687}.

\bibitem{Mohapatra:1979ia}
R.~N. Mohapatra and G.~Senjanovic, ``{Neutrino Mass and Spontaneous Parity Nonconservation},'' \href{http://dx.doi.org/10.1103/PhysRevLett.44.912}{{\em Phys. Rev. Lett.} {\bfseries 44} (1980) 912}.

\end{thebibliography}\endgroup
\end{document}